\documentstyle[sprocl]{article}

\input{psfig.sty}

\def\Journal#1#2#3#4{{#1} {\bf #2}, #3 (#4)}

\def\NPB{{\em Nucl. Phys.} B}
\def\PLB{{\em Phys. Lett.}  B}
\def\PRL{\em Phys. Rev. Lett.}
\def\PRD{{\em Phys. Rev.} D}
\def\ZPC{{\em Z. Phys.} C}

\def\CPC{\em Computer Phys. Commun.}

\def\ra{\rightarrow}

\def\be{\begin{equation}}
\def\ee{\end{equation}}
\def\bea{\begin{eqnarray}}
\def\eea{\end{eqnarray}}

\def\ra{\rightarrow \,\,}

\def\jpsi{$J/\psi$ \,\,}

\begin{document}

\title{HARD AND SOFT QCD IN CHARMONIUM PRODUCTION${}^\dag$\footnote[0]
{${}^\dag$Presented at the HADRON PHYSICS 2000 Workshop, Caraguatatuba, 
Brazil, April 2000}}

\author{{\underline {C.B. MARIOTTO}}$^{1,2}$,
M.B. GAY DUCATI$^{1}$, G. INGELMAN$^{2}$}
\address{$^{1}$ Instituto de F\'{\i}sica, 
Univ. Federal do Rio Grande do Sul,
\\ 
Box 15051, 91501-960 Porto Alegre, Brazil;
E-mail: mariotto@if.ufrgs.br}
\address{$^{2}$ High Energy Physics, Uppsala University,
Box 535, S-751~21 Uppsala, Sweden}


\maketitle\abstracts{We show that both hard perturbative and soft
non-perturbative QCD effects are important in hadroproduction
of charmonium. Observed $x_F$ and $p_\perp$ distributions of $J/\psi$
in fixed target experiments can be well described by our Monte Carlo model
based on perturbative QCD and the Color Evaporation Model. }

\vspace{-0.3cm}

Aiming at an understanding of non-perturbative QCD (non-pQCD),
we study effects of soft color interactions on charmonium production.
This is a particularly interesting case, since the charm quark mass $m_c$
is a large enough scale for perturbative QCD (pQCD) to be used as a basis,
but small enough for non-perturbative effects to be of importance.
This provides an interesting interplay between hard and soft QCD dynamics.

In the Color Singlet Model~\cite{CSM} (CSM) a colour singlet $c\bar{c}$ pair is
produced in a conventional pQCD process. However, this model gives a cross
section more than one order of magnitude lower than the observed one for
high-$p_\perp$ charmonium at the Tevatron~\cite{high-pt-data}. This
surprisingly high rate can be reproduced well by the Color Evaporation Model
(CEM)~\cite{HALZENquantit} and the Soft Color Interaction~\cite{SCI} model
which not only include conventional pQCD but also aspects of non-pQCD 
dynamics. In essence, a 
perturbatively produced 
$c\bar{c}$ pair in a color octet state
can be transformed into a color singlet state through soft gluon exchange.

In CEM a simple statistical 
color factor $1/9$ gives the probability that a
$c\bar{c}$ pair is in a singlet state. For pairs with mass below the threshold
for open charm production, $m_{c\bar{c}}<2m_D$, the fraction giving a $J/\psi$
is given by an additional non-perturbative parameter $\rho_{J/\psi}=0.5$.
We have earlier found~\cite{CEMnosso} that CEM can replace the Pomeron-based
model~\cite{Rys} to describe photoproduction of $J/\psi$ and $D$.
As a result of our further investigation of the CEM model, we here present
preliminary results for the $x_F$ and $p_\perp$ distributions of $J/\psi$ in
hadroproduction.

\begin{figure}[tp]
\begin{center}
\vspace*{-4mm}
\centerline{\psfig{file=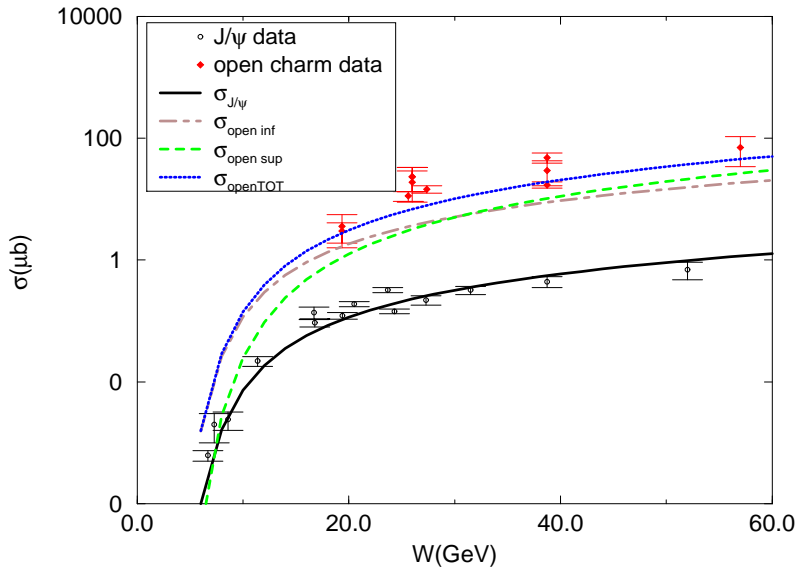,width=112mm}} 
\end{center}
\vspace*{-12mm}
\caption{Total cross section of $J/\psi$ and 
open charm ($D$) in hadroproduction as obtained
with CEM based on the LO pQCD subprocesses and including a factor $K=3.3$. The
curves $\sigma _{open\; sup}$ and $\sigma _{open\; inf}$ are the contributions 
to open charm for $m_{c\bar{c}}$ above and below $2m_D$, the latter for
$c\bar{c}$ in a color octet state.}
\vspace*{-4mm}
\label{fig1}
\end{figure}

The total cross section for \jpsi in hadroproduction was calculated analytically
using leading order (LO) QCD, where the contributing processes are
$gg \rightarrow c\overline{c}$ and $q\overline{q} \rightarrow c\overline{c}$.
Our results in fig.~\ref{fig1} show an energy
dependence in agreement with data. However, 
an overall normalisation factor $K=3.3$ was needed in order to describe 
the data. This may be attributed to either higher order pQCD
corrections that were not included in our calculation, or to some soft dynamics
that could also play an important role.

In order to understand the origin of this $K$-factor, we studied also some
important differential distributions, namely $x_F$ (longitudinal momentum
fraction) and $p_\perp$ of $J/\psi$ in fixed target experiments.
For this purpose we have added the Color Evaporation Model to the
{\sc Pythia}~\cite{Pythia} Monte Carlo program so that complete events can be
simulated. The LO pQCD processes are the same as described above with
the charm mass included in the matrix element. The importance of higher order
pQCD contributions 
was estimated by the initial and final state parton showers
where a $c\bar{c}$ pair can be produced. In this case we include all
$2 \ra 2$ processes ($gg \ra gg$, $qg \ra qg$, $qq \ra qq$ etc) as hard
scattering basis, except those producing a $c\bar{c}$ to avoid double counting.

\begin{figure}[t]
\begin{center}
\vspace*{-4mm}
\centerline{\psfig{file=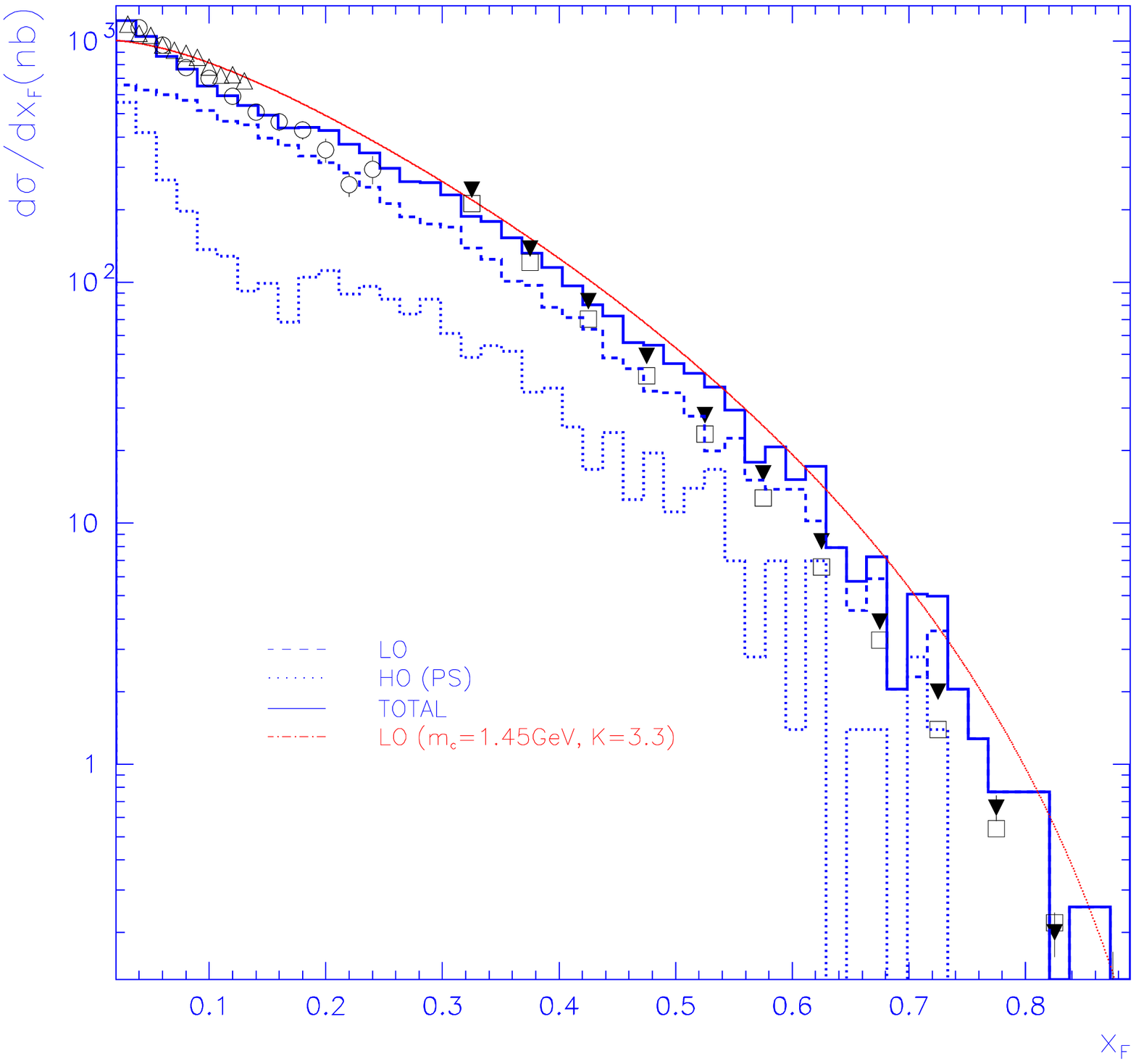,width=65mm}}
\vspace*{-3mm}
\centerline{\psfig{file=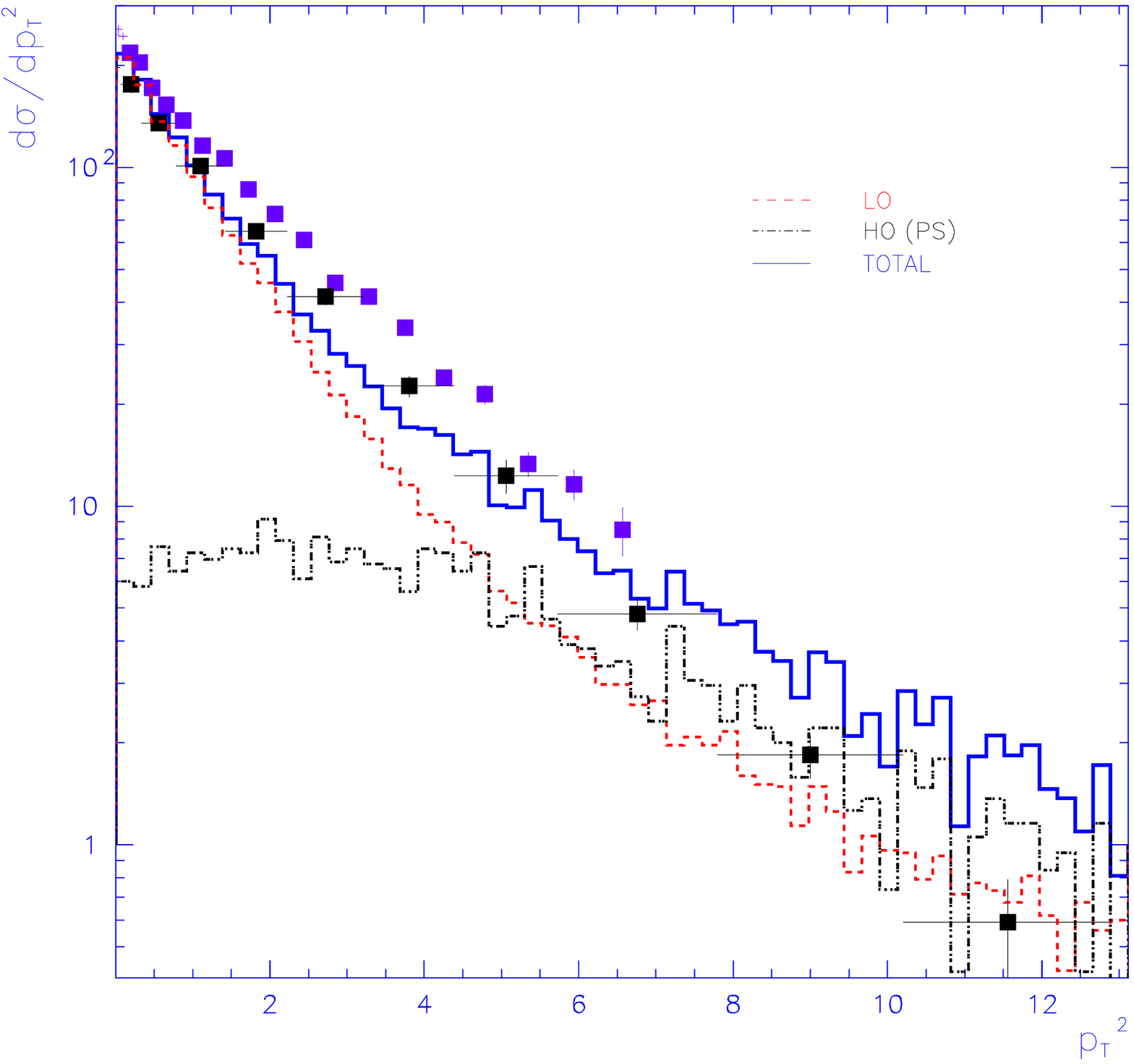,width=65mm}}
\vspace*{-3mm}
\caption{Distributions in $x_F$ and $p_\perp^2$ of $J/\psi$ in fixed target
hadroproduction with $E_{beam}=800\, GeV$. Data~\protect\cite{data} compared
to CEM results based on leading order (LO) matrix elements and higher orders
(HO) estimated by parton showers, and their sum (full histogram).
Also shown for $x_F$ (thin full curve) is the LO result with $m_c=1.45\, GeV$ 
and $K=3.3$.}
\label{xfbptdistr}
\end{center}
\vspace*{-10mm}
\end{figure}

Our results are shown in fig.~\ref{xfbptdistr}, 
where the $x_F$ and $p_\perp$ distributions of $J/\psi$ are compared to
fixed target data~\cite{data}.
With a charm mass $m_c=1.3\,GeV$, there is no need for a $K$ factor,
provided the higher order pQCD contributions and the soft color effects
are included. Thus, both higher order pQCD and soft effects are important.

The higher order pQCD gives important contributions at small $x_F$ and
large $p_\perp$. The distribution in $p_\perp$ gives further insights
on the interplay between hard and soft processes. The data can be reasonably
well described by a gaussian of width $1\, GeV$, which suggests that both
hard and soft contributions are present. The model has three basic sources
giving tranverse momentum to the $J/\psi$.
The first is from the pQCD processes described by
the $2\to 2$ matrix elements and by the parton showers. 
The LO $c\bar{c}$ process dominates strongly the $J/\psi$ production at 
low $p_\perp$, but not at larger $p_\perp$ because a larger momentum transfer
in the $2\to 2$ process leads to $M_{c\bar{c}}>2m_D$ and open charm production.
Instead, the higher order pQCD processes, where typically a gluon from the 
LO $2\to 2$ process splits into a $c\bar{c}$ pair, 
is important for the $J/\psi$ production at larger
$p_\perp$ as can be seen in fig.~\ref{xfbptdistr}b.

\nopagebreak
The two other sources of transverse momentum are due to soft dynamics.
One is the Fermi motion of partons inside the initial state hadrons,
for which \nolinebreak we use the conventional gaussian distribution with
$\langle k^2_T \rangle = (0.59\,GeV/c)^2$ used in {\sc Pythia}.
The other source is the soft momenta that may be associated with the soft
gluon exchange that neutralize color. We model this source by giving the
$c\bar{c}$ pair an additional $p_\perp$ in random direction 
using the same gaussian distribution as for the soft Fermi motion. 
This results in a reasonable description of the data as shown in fig.~\ref{xfbptdistr}b.
Thus, both hard pQCD and soft non-pQCD effects should be considered 
in order to understand the observed $p_\perp$ distribution.

In conclusion, we have found that both higher order pQCD and soft non-pQCD
effects are important for the understanding of charmonium production.
The soft interactions can be successfully modelled, e.g.\ by the Color
Evaporation Model which we have made more dynamically explicit by implementing
it in the {\sc Pythia} Monte Carlo. Simulated events then give access to more
detailed information on $J/\psi$ production.
Our results, which depend on a few parameters related to basic unknown
features of non-pQCD, provide a reasonable description of the overall
cross section of $J/\psi$ in hadroproduction, without the use of an arbitrary
$K$-factor.
The differential distributions in $x_F$ and $p_\perp$ of the \jpsi are also
well reproduced.

\vspace*{2mm}
This work was partially financed by CNPq, by Programa de Apoio a N\'ucleos de
Excel\^encia (PRONEX), BRAZIL, and by the
Swedish Natural Science Research Council.
C.B.M.\ would like to thank Uppsala University for the
kind hospitality.

\end{document}